\documentclass[prb,preprint,groupedaddress,showpacs,preprintnumbers,amsmath,amssymb]{revtex4}

\usepackage{graphicx}
\usepackage{dcolumn}
\usepackage{bm}

\begin{document}


\title{Dynamics of a two-level system coupled with a quantum oscillator: The very strong 
coupling limit}

\author{Titus Sandu }
 \email{titus.sandu@umontreal.ca}
\affiliation{D\'{e}partement de chimie, Universit\'{e} de Montr\'{e}al, 
C.P. 6128, succursale Centre-ville, Montr\'{e}al, Qu\'{e}bec H3C 3J7, Canada
 }%


\author{}
\affiliation{}


\date{\today}

\begin{abstract}
The time-dependent behavior of a two-level system interacting with a quantum 
oscillator system is analyzed 
in the case of a coupling larger than both 
the energy separation between the two levels and the energy 
of quantum oscillator ($\Omega < \omega <  \lambda $, where 
$\Omega $ is the frequency of the transition between the two levels, $\omega 
$ is the frequency of the oscillator, and $\lambda $ is the coupling between 
the two-level system and the oscillator). Our calculations 
show that the amplitude of the 
expectation value of the oscillator coordinate decreases as the two-level system undergoes the 
transition from one level to the other, while the transfer probability 
between the levels is staircase-like. This behavior is explained by 
the interplay between the adiabatic and the non-adiabatic regimes 
encountered during the dynamics with the system acting as a quantum counterpart of the Landau-Zener model. The transition between the two levels occurs as long as the expectation 
value of the oscillator coordinate is driven close to zero. On the contrary, if the 
initial conditions are set such that the expectation values of the oscillator coordinate are 
far from zero, the system will remain locked on one level.  
\end{abstract}
\pacs{42.50.Md, 42.50.Hz,63.20.Kr, 85.25.Cp, 85.85.+j}

\maketitle



One of the most studied quantum-mechanical models is the two-level system
interacting with a quantum oscillator. 
It is used in a wide range of phenomena, especially in 
atomic physics where it describes a
two-level atom coupled to a quantized electromagnetic 
field.\cite{Scully96,Raimond01}  
A challenging enterprise is to extend the model to "artificial atoms" in 
a condensed-matter environment. These small solid-state devices like flux 
lines threading a superconducting loop, charges in Cooper pair boxes, and single-electron spins 
exhibit quantum-mechanical properties which can be manipulated by currents 
and voltages.\cite{Vion02,Yu02,Golovach02} 

The solid-state devices offer wider regimes for the coupling strength between the two-level system 
and the oscillator. Typically, the coupling strength in atomic systems is 
$\lambda \mathord{\left/ {\vphantom {\lambda \omega }} \right. 
\kern-\nulldelimiterspace} \omega = 10^{ - 7} - 
10^{ - 6}$.\cite{Raimond01} Similar 
dipolar coupling in Cooper pair boxes and Josephson charge qubits is $3-4$ order of magnitude larger than 
the coupling in atomic systems.\cite{Wallraff04,Wallraff05} In contrast to the dipolar coupling, the 
capacitive and inductive couplings 
show even larger coupling strengths.\cite{Chiorescu04,Armour02,Irish03} Recently it has been argued that values 
$\lambda \mathord{\left/ {\vphantom {\lambda \omega }} \right. 
\kern-\nulldelimiterspace} \omega \approx 1$ are possible 
to be achieved experimentally.\cite{Irish05} In this brief report we examine the regime 
$\lambda \mathord{\left/ {\vphantom {\lambda \omega }} \right. 
\kern-\nulldelimiterspace} \omega > 1$. We will show that, although the oscillator dynamics
 "follows" the dynamics of the two-level system as in the case studied in Ref.~\onlinecite{Irish05}, 
the general features are 
different: the system undergoes the transition from one level to the other with a sudden change in 
the transition probability, whenever the expectation value of 
the oscillator coordinate is close to zero.

The Hamiltonian of the system is written as 
($\hbar = 1)$:

\begin{equation}
\label{eq:ham}
H = \frac{p^2 + \omega ^2q^2}{2} + \lambda q\sigma _z + \Omega 
\sigma _x ,
\end{equation}

\noindent
where $\sigma _{z }$, and $\sigma _{x }$ are the spin 
operator matrices, and $p$ and $q$ are the oscillator coordinates. The essential 
parameters are \textit{$\omega $}, \textit{$\lambda $}, and \textit{$\Omega $}, 
associated with the frequency of the oscillator, the 
coupling strength of the two-level system with the oscillator, and the splitting 
frequency of two-level system, respectively (our parameter $\lambda$ is 
scaled up by a factor of $2\sqrt{2}$ with respect to parameter $\lambda$ 
in Ref.~\onlinecite{Irish05}). In 
Ref.~\onlinecite{Irish05} it was assumed that the splitting frequency is much 
smaller than the frequency of the oscillator, thus the problem can be cast 
into the displaced oscillator basis which is the basis for the first two 
terms of the Hamiltonian (\ref{eq:ham}). This displaced oscillator basis is 
found by applying the unitary transformation $U = 
\exp \left( {\textstyle{{i\,\lambda \,p} \over {\omega^2 }}\sigma _z } 
\right)$ to the Hamiltonian (\ref{eq:ham}). The transformed Hamiltonian becomes\cite{Wagner79}

\begin{equation}
\label{eq:ham1}
\begin{array}{l}
 H' = \frac{p^2 + \omega ^2q^2}{2} + \frac{\Omega}{2} \left[ {\sigma ^ + 
\exp \left( { - i\frac{\lambda \,p}{\omega^2 }} \right) + h.c.} \right] - \frac{1}{8}
\frac{\lambda^2 }{\omega^2 }  \\ 
 \end{array},
\end{equation}

\noindent
with $\sigma ^ + $($\sigma ^ - $) as the spin-$\frac{1}{2}$ creation (annihilation) operator, 
than time-dependent or/and time-independent perturbation calculations can be 
performed as long as $\Omega \mathord{\left/ {\vphantom {\Omega \omega }} 
\right. \kern-\nulldelimiterspace} \omega < < 1$. 
A similar path was followed by Schweber. \cite{Schweber67} 
Perturbation calculations 
on the energy spectrum have been performed and compared with the exact calculations 
in Fig. 3 of Ref.~\onlinecite{Irish05}. The authors extended the comparison to the 
regime $\Omega \mathord{\left/ {\vphantom {\Omega \omega }} \right. 
\kern-\nulldelimiterspace} \omega \ge 1$ and they found some quantitative and 
qualitative resemblance between the exact solution and their approximate 
solution (Figs. 3(b) and 3(d) in their paper). The quantitative and qualitative 
agreement shown in Fig. 3 of Ref.~\onlinecite{Irish05} can be explained in simple terms as 
follows. The net effect of 
$\exp \left( {\textstyle{{i\,\lambda \,p} \over {2\omega^2 }}}\right)$  
on the wave function is to displace it by the amount $\frac{\lambda} {2\omega^2 }$. 
Thus, the effective splitting given by the second term of the Hamiltonian (\ref{eq:ham1}) 
will be quenched, such that 
perturbation calculations on the Hamiltonian (\ref{eq:ham1}) can be extended to larger values 
of $\Omega$.  

Irish and coworkers\cite{Irish05} studied 
the collapse and revival of the wave function for $\Omega \mathord{\left/ {\vphantom {\Omega \omega }} 
\right. \kern-\nulldelimiterspace} \omega < < 1$ and $\lambda \mathord{\left/ {\vphantom {\lambda \omega }} \right. 
\kern-\nulldelimiterspace} \omega < 1$.
In the following we will explore the dynamics of the very strong coupling regime 
($\lambda \mathord{\left/ {\vphantom {\lambda \omega }} \right. 
\kern-\nulldelimiterspace} \omega > 1$ and $\Omega \mathord{\left/ {\vphantom {\Omega \omega }} 
\right. \kern-\nulldelimiterspace} \omega <  1$). The Hamiltonian (\ref{eq:ham1}) would be able, in principle, 
to explain the 
dynamics in the very strong regime because the effective coupling between the two wells generated by 
the displaced oscillators 
is quenched by the 
separation of these two potential wells. The effective coupling between the two wells decreases 
exponentially \cite{Irish05} with 
the strength of $\lambda$ as it can be shown also by analyzing 
Eq.~(\ref{eq:ham1}). However, to gain a new insight into the dynamics of the very strong 
regime, we follow a different approach. 
We perform another unitary transformation,\cite{Wagner79,Sandu03}  
$U = \exp \left( {i\Lambda \left( q \right)\,\sigma _y } \right)$, on the Hamiltonian (\ref{eq:ham}), 
with $\tan \left( {\Lambda \left( q \right)} \right) = - \frac{\Omega 
}{\lambda q}$, to obtain adiabatic motions which are valid for either 
very strong coupling $\lambda > > \Omega ,\omega $ (case A) or 
large level splitting $\Omega > > \lambda ,\omega $ (case B). 
The intuitive picture of the transformation is a $q$-dependent rotation around 
the $y$-axis that brings the effective field seen by the two-level system along the $z$-axis.
The Hamiltonian 
resulting from the above unitary transformation is

\begin{equation}
\label{eq:ham2}
\begin{array}{l}
 H' = \frac{1}{2}\left( {p^2 + \omega ^2q^2} \right) + 
\frac{1}{8}\frac{\Omega ^2\lambda ^2 }{\left( {\Omega ^2 + \lambda 
^2 q^2} \right)^2} + \sigma _z \sqrt {\Omega 
^2 + \lambda ^2 q^2} + \\ 
 \quad \quad + \frac{\lambda \Omega }{2}\sigma _y \left( 
{p\frac{1}{\Omega ^2 + \lambda ^2 q^2} + 
\frac{1}{\Omega ^2 + \lambda ^2 q^2}p} \right). 
\\ 
 \end{array}
\end{equation}

\begin{figure}
\includegraphics{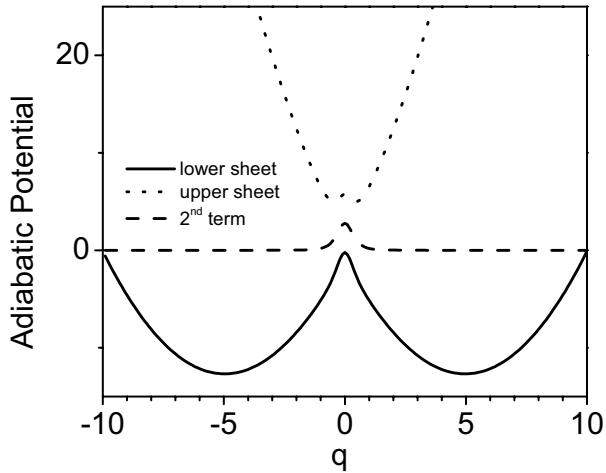}
\caption{\label{fig:1} The shape of the two sheets of the adiabatic 
potential generated by the Hamiltonian 
in Eq.~(\ref{eq:hamad}). The solid line is the lower sheet and the 
dotted line is the upper sheet. The second term in Eq.~(\ref{eq:hamad}) is 
sharply peaked around origin and it is plotted with dashed line.}
\end{figure}

\noindent
The above Hamiltonian (\ref{eq:ham2}) will be used below for our investigation. 
The last term ($\sigma _y $-term) in Eq.~(\ref{eq:ham2}) is small as long as 
$\lambda q > > \Omega$ (that is supposed to be fulfilled in Case A) or 
$\lambda q < < \Omega$ (that is fulfilled in Case B). The inequalities 
$\lambda q > > \Omega$ and $\lambda q < < \Omega$ should be understood as operator inequalities 
in the sense that they have to be satisfied as inequalities for matrix elements in a certain basis.
The $\sigma _y $-term is the non-diagonal term 
and it accounts for the non-adiabaticity. 
Without the $\sigma _y $-term, the Hamiltonian (\ref{eq:ham2}) reveals 
an adiabatic motion with one part (either the two-level system or the oscillator) becoming fast, while the other 
part becoming slow.\cite{Sandu03} 
Thus, the adiabatic motion is 
generated by the first three terms,

\begin{equation}
\label{eq:hamad}
H'_{ad} = \frac{1}{2}\left( {p^2 + \omega ^2q^2} \right) + 
\frac{1}{8}\frac{\Omega ^2\lambda ^2 }{\left( {\Omega ^2 + \lambda 
^2 q^2} \right)^2} + \sigma _z \sqrt {\Omega 
^2 + \lambda ^2 q^2} .
\end{equation}

\begin{figure*}
\includegraphics{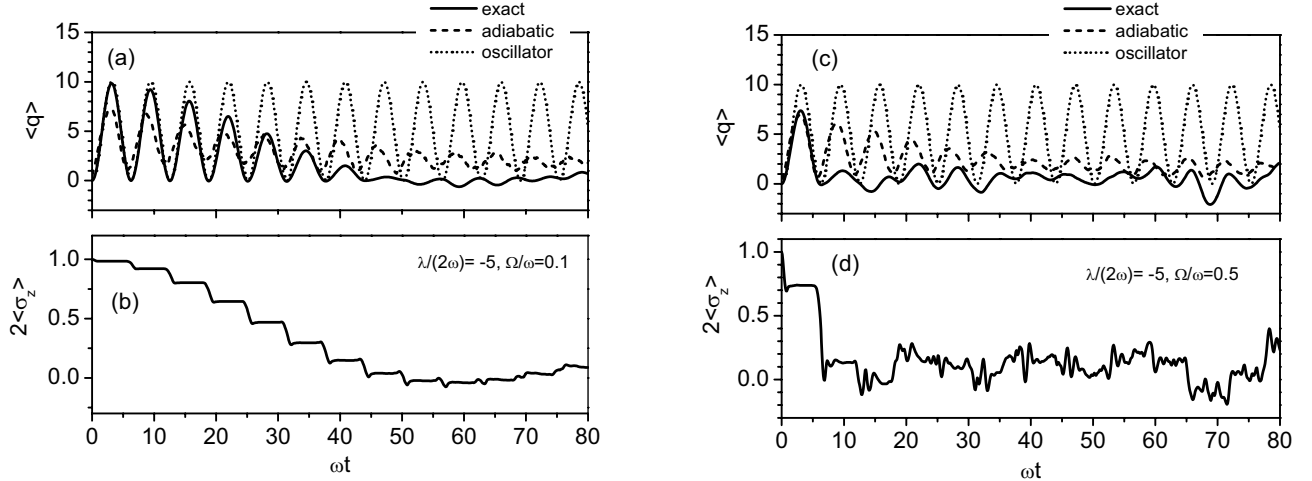}
\caption{\label{fig:2}  Time-dependent behavior of the expectation value of: (a) $q-$coordinate, with 
${\Omega \over \omega} = 0.1$ and ${\lambda \over 2\omega} = -5$; (b) $2\sigma_z$, with 
${\Omega \over \omega} = 0.1$ and ${\lambda \over 2\omega} = -5$; (c) $q-$coordinate, with 
${\Omega \over \omega} = 0.5$ and ${\lambda \over 2\omega} = -5$; (d) $2\sigma_z$, with 
${\Omega \over \omega} = 0.5$ and ${\lambda \over 2\omega} = -5$. The solid line denotes the exact dynamics, 
the dashed line represents the adiabatic dynamics, and the dotted line shows the corresponding 
displaced oscillator. For convenience, we have chosen negative values of $\lambda$. We notice that the 
strength of the coupling is, actually, $\frac {\lambda}{2}$.}
\end{figure*}

In the very strong coupling case ($\lambda > > \Omega ,\omega $), there are two adiabatic potential 
sheets coupled by the $\sigma _y $ term.\cite{Wagner79,Sandu03} The shape of the adiabatic 
potentials is presented in Fig.~\ref{fig:1}. Dynamically, the oscillator is 
fast and the two-level system is slow.\cite{Sandu03} The lower adiabatic sheet has two 
minima located at

\begin{equation}
\label{eq:qmin}
q_{\min } = \mp \sqrt {\frac{\lambda ^2}{4\omega ^4} - \frac{\Omega 
^2}{\lambda ^2 }}, 
\end{equation}
with the value 

\begin{equation}
\label{eq:Vqmin}
V\left( {q_{\min } } \right) = - \frac{\omega^2 \Omega ^2}{2\lambda ^2} - 
\frac{\lambda ^2}{8\omega^2},
\end{equation}

and its second derivative at the minimum points 

\begin{equation}
\label{eq:omega}
\omega _{\min }^2 = \omega ^2\left( {1 - \frac{4\Omega ^2\omega ^4}{\lambda 
^4}} \right).
\end{equation}

\noindent
The lower adiabatic sheet is very close to 
the unperturbed ($\Omega = 0$) displaced harmonic potential 
sheets as long as $\lambda > > \Omega ,\omega $. Therefore one might expect similar 
dynamic behavior 
as the one studied in Ref.~\onlinecite{Irish05}. 
However, the dynamics is different for very strong coupling. In Fig.~\ref{fig:2} we show the dynamics 
of the expectation values of $q$-coordinate and $\sigma _z$. The details of numerical integration 
are given in Ref.~\onlinecite{Sandu03}.
We compare 
the exact dynamics with the dynamics of the adiabatic potential (Eq.~(\ref{eq:hamad})), and 
with the dynamics of a displaced oscillator ($\Omega = 0$). One can notice in 
Fig.~\ref{fig:2} that the amplitude of the expectation value of $q$-coordinate 
decreases with time. It occurs as the system undergoes the transition from one 
level to another. In the same time, the transfer probability is staircase-like. 
Moreover, the slope of the amplitude decrease depends on $\Omega $ but the 
frequency of the oscillations does not change significantly.
Although it is weaker, the same quenching of $q$-coordinate appears 
for the adiabatic motion generated by Eq.~(\ref{eq:hamad}). It is weakly dependent on
$\Omega$ and it occurs as the system undergoes 
the transition from one potential well to the other potential well. 
This was pointed out by Wagner\cite{Wagner79} who showed that the transition rate from one 
well to the other is dependent on $\Omega$ in second order in the adiabatic approximation. 
In contrast to his paper\cite{Wagner79}, Fig.~\ref{fig:2} shows clearly that the adiabatic 
motion is not an accurate description of the full dynamics.  
This behavior will be explained 
below as interplay between the adiabatic and the 
non-adiabatic regimes encountered during the dynamics. 

In order to explain the dynamics, we employ the equation of motion 
in the Heisenberg picture for an operator, $\frac{d}{dt}A = i\left[ {H,A} \right]$. 
Taking the expectation value, the equation becomes $\frac{d}{dt}\langle A\rangle = i\langle \left[ {H,A} \right]\rangle $. 
We apply this last equation to the Hamiltonian represented by Eq.~(\ref{eq:ham}). The corresponding equations are

\begin{equation}
\label{eq:Heisenberg}
 \begin{array}{l}
 \frac{d}{dt}\langle q\rangle = \langle p\rangle , \\ 
 \frac{d}{dt}\langle p\rangle = - \omega^2 \langle q\rangle - \lambda \langle \sigma _z \rangle ,\\ 
 \frac{d}{dt}\langle \sigma _z \rangle =  \Omega \langle \sigma _y \rangle , \\ 
 \frac{d}{dt}\langle \sigma _x \rangle = - \lambda \langle q\sigma _y \rangle , \\ 
 \frac{d}{dt}\langle \sigma _y \rangle =  \lambda \langle q\sigma _x \rangle - \Omega \langle \sigma _z \rangle , \\ 
 \ldots \ldots \ldots \ldots \ldots \ldots \ldots \ldots \\ 
 \end{array} 
\end{equation}

\noindent
Eq.~(\ref{eq:Heisenberg}) is an infinite chain of coupled ordinary differential equations. The chain can be broken by 
making assumptions like

\begin{equation}
\label{eq:adiabatic}
\begin{array}{l}
 \langle q\sigma _y \rangle \cong \langle q\rangle \langle \sigma _y \rangle 
, \\ 
 \langle q\sigma _x \rangle \cong \langle q\rangle \langle \sigma _x \rangle 
. \\ 
 \end{array}
\end{equation}

\noindent
The approximations made in Eq.~(\ref{eq:adiabatic}) are valid if one part (the oscillator) is 
fast and the other part (the two-level system) is slow as in the usual adiabatic approximation.\cite{Graham84,Bersuker89} Thus, 
combining (\ref{eq:Heisenberg}) 
and (\ref{eq:adiabatic}), one can show that Eq.~(\ref{eq:Heisenberg}) can be approximated 
and then recast as

\begin{equation}
\label{eq:dimer}
\begin{array}{l}
 \frac{d}{dt}\langle q\rangle = \langle p\rangle , \\ 
 \frac{d}{dt}\langle p\rangle = - \omega^2 \langle q\rangle - \frac{\lambda}{2} \mbox{(}|f_1|^2 
- |f_2|^2\mbox{)} ,\\
 i\mathop \frac{d}{dt} f _1  = \frac{\lambda }{2}\left\langle q 
      \right\rangle f _1 + \frac{\Omega }{2} f _2 \\
 i\mathop \frac{d}{dt} f _2 = - \frac{\lambda }{2}\left\langle q \right\rangle f _2 + 
\frac{\Omega }{2} f _1, \\ 
 \end{array}
\end{equation}

\noindent
with $f_1$ ($f_2$) as being the probability function of the level 1(2). In other 
words, $|f_1|^2$($|f_2|^2$) is the probability of the system to be on the level 1(2). 

\begin{figure*}
\includegraphics{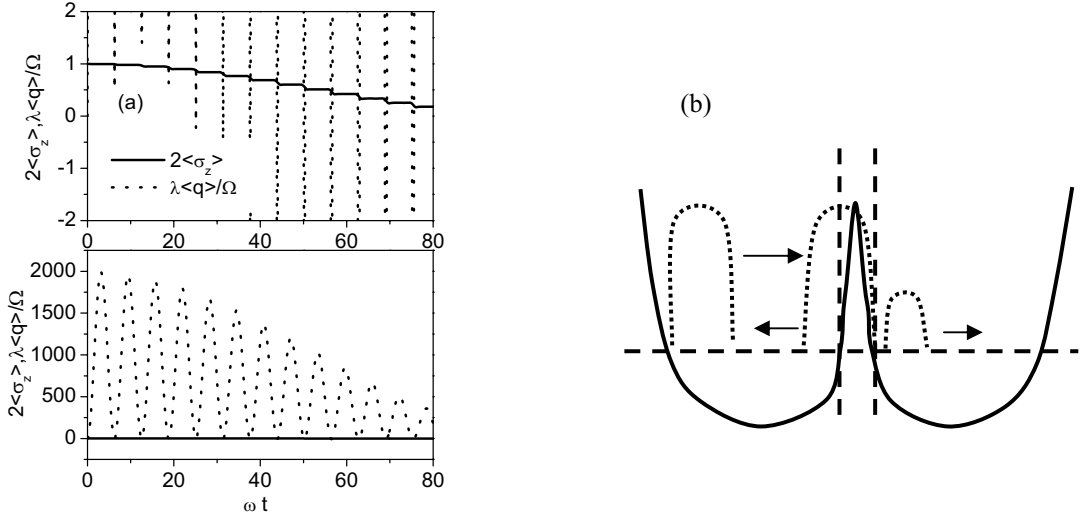}
\caption{\label{fig:3} (a) Comparison between the dynamics of $\langle \sigma_z \rangle$ (solid line) and 
$\lambda \langle q \rangle / \Omega$ (dotted line). It shows that $\langle \sigma_z \rangle$ changes whenever 
$\lambda \langle q \rangle / \Omega$ is close to $0$. The upper panel shows the scale around origin, while the 
lower panel shows the full scale of the plots. (b) Pictorial explanation of the dynamics of the 
two-level system in this very strong coupling case. The transition from the left well to the right well occurs 
at the origin where the adiabatic condition ($\lambda \langle q\rangle > > \Omega$) is not fulfilled.}
\end{figure*}

Eq.~(\ref{eq:dimer}) sheds a better light on the relationship between the oscillator and the two-level system.  
The first two equations in (\ref{eq:dimer}) are the equations of the classical harmonic oscillator displaced by the 
amount $\frac{\lambda}{2} \mbox{(}|f_1|^2 - |f_2|^2\mbox{)}$.  This will imply that the time variation of 
$\mbox{(}|f_1|^2 - |f_2|^2\mbox{)}= 2\langle \sigma _z \rangle $ 
modulates the amplitude of the oscillator as it can be seen in Fig.~\ref{fig:2}. 
The last two equations 
in (\ref{eq:dimer}) explain the time-dependent behavior of $\mbox{(}|f_1|^2 - |f_2|^2\mbox{)}$. First, let us assume 
that $\lambda \langle q\rangle > > \Omega$, which is the adiabatic condition and it is supposed to be true most of 
the time for $\lambda > > \Omega ,\omega $. 
The last two equations in (\ref{eq:dimer}) will be approximated by
$i\mathop \frac{d}{dt} f _1  \cong  \frac{\lambda }{2}\left\langle q \right\rangle f _1$ and $i\mathop \frac{d}{dt} f _2 
\cong  - \frac{\lambda }{2}\left\langle q \right\rangle f _2$.  This means that the probability functions $f_1$ and $f_2$ 
acquire just a phase factor. Therefore, there is no mixing between $f_1$ and $f_2$, and 
$\langle \sigma_z \rangle $ is constant in time. Now, let us assume that the reverse is true, 
$\lambda \langle q\rangle <\Omega$. Then, the last two equations in (\ref{eq:dimer}) will be approximated by 
$i\mathop \frac{d}{dt} f _1  \cong \frac{\Omega }{2} f _2$ and $i\mathop \frac{d}{dt} f _2 \cong \frac{\Omega }{2} f _1$, 
i.e., $f_1$ and $f_2$ will mix, and $\langle \sigma _z \rangle$ will change. These assertions are proven numerically 
in Fig.~\ref{fig:3}(a), where one can see that $\langle \sigma_z \rangle$ changes whenever $\lambda \langle q\rangle/\Omega \cong 0$. 

In Fig.~\ref{fig:3}(b), we give 
a simple explanation of the dynamics of the transition from level $1$ to level $2$. Suppose that the wave packet of the 
system is in the left well which is associated with level $1$. As soon as the wave packet reaches regions close to 
the origin it "sheds" a part of itself into the right well and the rest of it returns into the left well. 
Basically, the transition 
from level $1$ to level $2$ occurs during short periods of time when the system is out of its adiabatic regime.
The picture presented here is closely related to Landau-Zener theory\cite{Landau32,Zener32} that treats a quantum two-level system placed 
in a slowly varying external field. Near the crossing point the adiabaticity is violated and the system can 
escape from the state it occupied initially to another one. In the present case, the coupling is purely 
quantum through the quantum oscillator, hence the oscillator explores all possible trajectories 
(in the sense of Feynman's path integral) and the ones that explore the region at the level crossing can induce a non-zero 
flipping probability for the two-level system.

\begin{figure}
\includegraphics{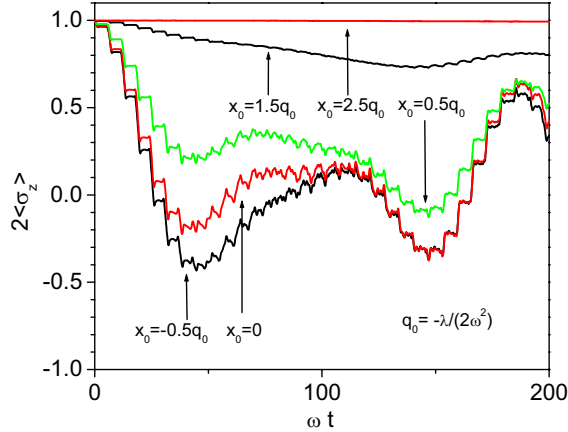}
\caption{\label{fig:4} (Color online) Dynamics of $\langle \sigma_z \rangle$ with different initial conditions.
The arrows indicate various initial conditions in terms of the displaced ground state wave-function 
of the initial oscillator. The oscillator displacements are $-0.5q_0$ (it produces the faster transition), 
$0$, $0.5q_0$, $1.5q_0$, and $2.5q_0$ (it locks the system on level 1),  
with $q_0 = -{ \lambda \over 2\omega^2}$. ${\Omega \over \omega} = 0.2$ and ${\lambda \over 2\omega} = 3$. }
\end{figure}

In Fig.~\ref{fig:4} we show the dynamics of $\langle \sigma_z \rangle$ as depending on initial conditions. We 
consider as initial 
conditions various displaced ground state wave-functions of the harmonic oscillator. The displacements are 
given in terms of $q_0 = -{ \lambda \over 2\omega^2}$, which is the displacement of the unperturbed oscillator 
($\Omega = 0$) corresponding to the level 1.
Assuming that the system is initially on the level 1, its $q$-coordinate, $\langle q \rangle$, will tend to 
oscillate around $q_0$, starting from the initial displacement. Fig.~\ref{fig:4} shows that the system undergoes 
faster transitions from level 1 to level 2 for those initial conditions which generate dynamics that 
fulfills the condition $\lambda \langle q \rangle << \Omega$ for longer time. In addition to that, the system 
can be locked on level 1 as long as the initial expectation value of $q$ is set to be 
close to $q _0$.

In conclusion, we investigated the dynamics of the two-level system interacting with a quantum harmonic oscillator 
in the regime of very strong coupling ($\lambda > \omega > \Omega) $). This regime generate an adiabatic 
motion defined by the condition $\lambda \langle q \rangle >> \Omega$. We have found that the amplitude of 
the expectation value of the oscillator coordinate, $\langle q \rangle$, varies similarly to the expectation 
value of $2\langle \sigma _z \rangle$ (the difference between the occupation probabilities of the 
two-level system). In addition, the difference $2\langle \sigma _z \rangle$ is stair-case like. This 
behavior is explained by the interplay between the adiabatic region 
(whenever $\lambda \langle q \rangle >> \Omega$) and non-adiabatic region 
(whenever $\lambda \langle q \rangle <\Omega$). The transition from one level to another occurs 
during short periods of time when the system is out of its adiabatic region. Thus, the system can 
be locked on one level if it is prepared to fulfill the adiabatic condition all the time.

\begin{acknowledgments}
The work has been supported in part by NSERC grants no. 311791-05 and 315160-05.
The author wish also to acknowledge the helpful discussions with dr. R. Iftimie.
\end{acknowledgments}




\end{document}